\documentclass[sigconf]{acmart}
\pdfoutput=1
\usepackage[utf8]{inputenc}

\usepackage{algorithmic}
\usepackage{cleveref}
\usepackage{comment}
\usepackage{lipsum}
\usepackage{graphicx}
\usepackage{siunitx}
\usepackage{subfigure}
\usepackage{textcomp}
\usepackage{todonotes}
\usepackage{xcolor}
\usepackage{xspace}

\theoremstyle{definition}
\newtheorem{researchquestion}{RQ}
\crefname{researchquestion}{RQ}{RQs}

\def\FOSS/{FOSS}
\def\SWH/{Software Heritage}
\def\DATALastCommitDate/{2021-07-07}

\newif\iflong
\longfalse

\newcommand{\CommitCount}{2.2 billion\xspace}
\newcommand{\AuthorCount}{43 million\xspace}
\newcommand{\ProjectCount}{160 million\xspace}

\def\DATAAuthorsRaw/{\num{43 381 366}}
\def\DATAAuthorsRawApprox/{43\,M}
\def\DATACommitsRaw/{\num{2 198 808 389}}
\def\DATACommitsPlausible/{\num{1 735 130 408}}
\def\DATATotalCommitsInSH/{about \ProjectCount}
\def\DATALastCommitDate/{2021-07-07}
\def\ALGtzname/{tz\&name}
\def\ALGemail/{ccTLD}
\def\DATACommitsWithEmail/{13\%}
\def\DATACommitsWithoutEmail/{87\%}
\def\DATACommitsTZZTwoThousand/{96\%}
\def\DATACommitsTZZTwoThousandTen/{64\%}
\def\DATACommitsTZZTwoThousandTwenty/{22\%}
\def\DATAWorldRegions/{12}  \def\PKGGG/{\mbox{\scshape\small gender-guesser}}
\def\TZANALYZED/{20}
\def\DATAZonedCommitsRatio/{70\%}
\def\DATAYearRange/{1971--2021}

\def\DATAAuthorsRmNondecodable/{\num{4 127}}
\def\DATAAuthorsRmNonletter/{\num{9 915 884}}
\def\DATAAuthorsRmEmail/{\num{84 954}}
\def\DATAAuthorsRmBlank/{\num{25 055}}
\def\DATAAuthorsRmToolong/{\num{46}}
\def\DATAAuthorsPlausible/{\num{33 351 300}}
\def\DATAAuthorsPlausibleApprox/{33\,M}
\def\DATAAuthorsPlausiblePct/{77\%}

\def\DATAKnownAuthorsPct/{64.1\%}
\def\DATAUnknownAuthorsPct/{35.9\%}
\def\DATAKnownAuthorsApprox/{21.4\,M}
\def\DATAKnownCommitsPct/{50.3\%}
\def\DATAUnknownCommitsPct/{49.7\%}
\def\DATAKnownCommitsApprox/{872.3\,M}

\def\DATAMaleAuthorsPct/{86.5\%}
\def\DATAMaleAuthorsApprox/{10.4\,M}
\def\DATAFemaleAuthorsPct/{13.5\%}
\def\DATAFemaleAuthorsApprox/{1.6\,M}

\def\DATAMaleCommitsPct/{91.9\%}
\def\DATAMaleCommitsApprox/{801.8\,M}
\def\DATAFemaleCommitsPct/{8.1\%}
\def\DATAFemaleCommitsApprox/{70.5\,M}

 \copyrightyear{2022}
\acmYear{2022}
\setcopyright{acmlicensed}
\acmConference[MSR '22]{19th International Conference on Mining Software Repositories}{May 23--24, 2022}{Pittsburgh, PA, USA}
\acmBooktitle{19th International Conference on Mining Software Repositories (MSR '22), May 23--24, 2022, Pittsburgh, PA, USA}
\acmPrice{15.00}
\acmDOI{10.1145/3524842.3528471}
\acmISBN{978-1-4503-9303-4/22/05}

\title{Geographic Diversity in Public Code Contributions}
\subtitle{An Exploratory Large-Scale Study Over 50 Years}
\author{Davide Rossi}
\email{daviderossi@unibo.it}
\orcid{0000-0003-2269-8094}
\affiliation{\institution{University of Bologna}
  \city{Bologna}
  \country{Italy}
}

\author{Stefano Zacchiroli}
\email{stefano.zacchiroli@telecom-paris.fr}
\orcid{0000-0002-4576-136X}
\affiliation{\institution{LTCI, Télécom Paris, Institut Polytechnique de Paris}
  \city{Paris}
  \country{France}
}

\begin{abstract}

  We conduct an exploratory, large-scale, longitudinal study of 50 years of
  commits to publicly available version control system repositories, in
  order to characterize the geographic diversity of contributors to public code
  and its evolution over time. We analyze in total \CommitCount commits
  collected by \SWH/ from \ProjectCount projects and authored by \AuthorCount
  authors during the \DATAYearRange/ time period. 
  We geolocate developers to
  \DATAWorldRegions/ world regions derived from the United Nation geoscheme,
  using as signals email top-level domains,
  author names compared with names distributions around the world, and UTC offsets 
  mined from commit metadata.

  We find evidence of the early dominance of North America in open source
  software, later joined by Europe. After that period, the geographic diversity 
  in public code has been constantly increasing.
  We also identify relevant historical shifts related to
  the UNIX wars, the increase of coding literacy in Central and South Asia,
  and broader phenomena like colonialism and people movement across
  countries (immigration/emigration).

\end{abstract}


\begin{document}
\maketitle

\section{Introduction}
\label{sec:intro}

\emph{Gender diversity}, or more often its lack thereof, among participants to
software development activities has been thoroughly studied in recent years. In
particular, the presence of, effects of, and countermeasures for \emph{gender
  bias} in Free/Open Source Software (FOSS) have received a lot of attention
over the past decade~\cite{david2008fossdevs, qiu2010kdewomen,
  nafus2012patches, kuechler2012genderfoss, vasilescu2014gender,
  oneil2016debiansurvey, robles2016womeninfoss, terrell2017gender,
  zacchiroli2021gender}.  \emph{Geographic diversity} is on the other hand the
kind of diversity that stems from participants in some global activity coming
from different world regions and cultures.

Geographic diversity in FOSS has received relatively little attention in scholarly
works. In particular, while seminal survey-based and
point-in-time medium-scale studies of the geographic origins of FOSS
contributors exist~\cite{ghosh2005understanding, david2008fossdevs,
  barahona2008geodiversity, takhteyev2010ossgeography, robles2014surveydataset,
  wachs2021ossgeography}, large-scale longitudinal studies of the geographic
origin of FOSS contributors are still lacking. Such a quantitative
characterization would be useful to inform decisions related to global
development teams~\cite{herbsleb2007globalsweng} and hiring strategies in the
information technology (IT) market, as well as contribute factual information
to the debates on the economic impact and sociology of FOSS around the world.

\paragraph{Contributions}

With this work we contribute to close this gap by conducting \textbf{the first
  longitudinal study of the geographic origin of contributors to public code
  over 50 years.} Specifically, we provide a preliminary answer to the
following research question:
\begin{researchquestion}
  From which world regions do authors of publicly available commits come from
  and how has it changed over the past 50 years?
  \label{rq:geodiversity}
\end{researchquestion}
We use as dataset the \SWH/ archive~\cite{swhipres2017} and analyze from it
\CommitCount commits archived from \ProjectCount projects and authored by
\AuthorCount authors during the 1971--2021 time period. 
We geolocate developers to
\DATAWorldRegions/ world regions, using as signals email country code top-level domains (ccTLDs) and 
author (first/last) names compared with name distributions around the world, and UTC offsets 
mined from commit metadata.

We find evidence of the early dominance of North America in open source
software, later joined by Europe. After that period, the geographic diversity 
in public code has been constantly increasing.
We also identify relevant historical shifts
related to the end of the UNIX wars and the increase of coding literacy in
Central and South Asia, as well as of broader phenomena like colonialism and
people movement across countries (immigration/emigration).

\paragraph{Data availability.}

A replication package for this paper is available from Zenodo at
\url{https://doi.org/10.5281/zenodo.6390355}~\cite{replication-package}.

 \section{Related Work}
\label{sec:related}

Both early and recent works~\cite{ghosh2005understanding, david2008fossdevs,
  robles2014surveydataset, oneil2016debiansurvey} have characterized the
geography of Free/Open Source Software (FOSS) using \emph{developer surveys},
which provide high-quality answers but are limited in size (2-5\,K developers)
and can be biased by participant sampling.

In 2008 Barahona et al.~\cite{barahona2008geodiversity} conducted a seminal
large-scale (for the time) study on FOSS \emph{geography using mining software
  repositories (MSR) techniques}. They analyzed the origin of 1\,M contributors
using the SourceForge user database and mailing list archives over the
1999--2005 period, using as signals information similar to ours: email domains
and UTC offsets. 
The studied period (7 years) in~\cite{barahona2008geodiversity} is shorter than 
what is studied in the present paper (50 years) and the data sources are 
largely different; with that in mind, our results show a slightly larger quote of 
European v.~North American contributions.

Another empirical work from 2010 by Takhteyev and
Hilts~\cite{takhteyev2010ossgeography} harvested self-declared geographic
locations of GitHub accounts recursively following their connections,
collecting information for $\approx$\,70\,K GitHub users.  A very recent
work~\cite{wachs2021ossgeography} by Wachs et al.~has geolocated half a million
GitHub users, having contributed at least 100 commits each, and who
self-declare locations on their GitHub profiles. While the study is
point-in-time as of 2021, the authors compare their findings
against~\cite{barahona2008geodiversity, takhteyev2010ossgeography} to
characterize the evolution of FOSS geography over the time snapshots taken by
the three studies.

Compared with previous empirical works, our study is much larger scale---having
analyzed \AuthorCount authors of \CommitCount commits from \ProjectCount
projects---longitudinal over 50 years of public code contributions rather than
point in time, and also more fine-grained (with year-by-year granularity over
the observed period). Methodologically, our study relies on Version Control
System (VCS) commit data rather than platform-declared location information.

Other works---in particular the work by Daniel~\cite{daniel2013ossdiversity}
and, more recently, Rastogi et al.~\cite{rastogi2016geobias,
  rastogi2018geobias, prana2021geogenderdiversity}---have studied geographic
\emph{diversity and bias}, i.e., the extent to which the origin of FOSS
developers affect their collaborative coding activities.
In this work we characterized geographic diversity in public code for the first
time at this scale, both in terms of contributors and observation period. We do
not tackle the bias angle, but provide empirical data and findings that can be
leveraged to that end as future work.

\emph{Global software engineering}~\cite{herbsleb2007globalsweng} is the
sub-field of software engineering that has analyzed the challenges of scaling
developer collaboration globally, including the specific concern of how to deal
with geographic diversity~\cite{holmstrom2006globaldev, fraser2014eastwest}.
Decades later the present study provides evidence that can be used, in the
specific case of public code and at a very large scale, to verify which
promises of global software engineering have borne fruit.

 \section{Methodology}
\label{sec:method}

\newif\ifgrowthfig  \growthfigtrue
\ifgrowthfig
\begin{figure}
  \includegraphics[width=\columnwidth]{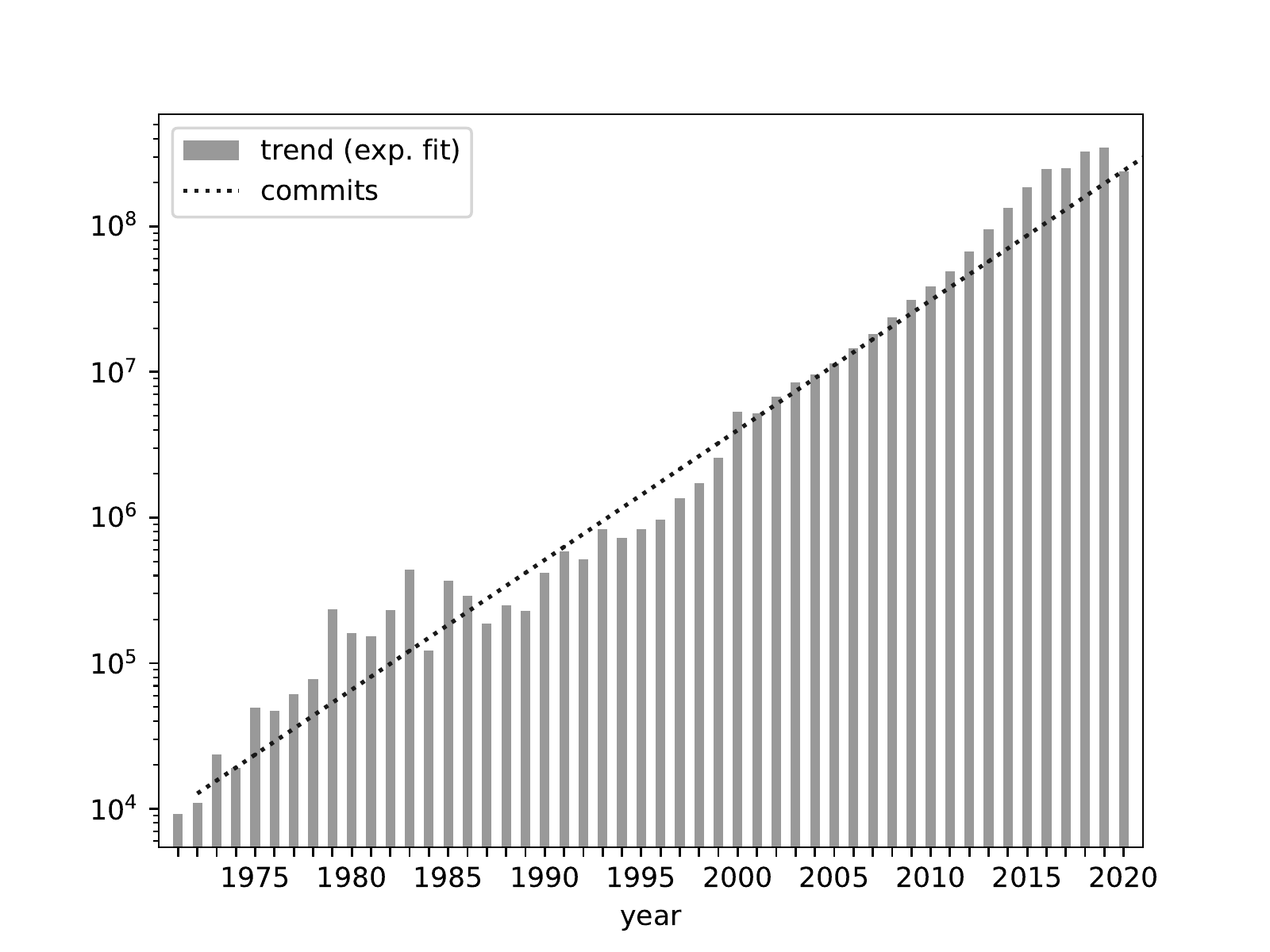}
  \caption{Yearly public commits over time (log scale).
}
  \label{fig:growth}
\end{figure}
\fi

\paragraph{Dataset}

We retrieved from \SWH/~\cite{swh-msr2019-dataset} all commits archived until \DATALastCommitDate/.
They amount to \DATACommitsRaw/ commits, unique by SHA1 identifier, harvested from \DATATotalCommitsInSH/ public projects coming from major development forges (GitHub, GitLab, etc.) and package repositories (Debian, PyPI, NPM, etc.).
Commits in the dataset are by \DATAAuthorsRaw/ authors, unique by $\langle$name, email$\rangle$ pairs.
The dataset came as two relational tables, one for commits and one for authors, with the former referencing the latter via a foreign key.
For each entry in the author table we have author full name and email as two separate strings of raw bytes.

We removed implausible or unusable names that: are not decodable as UTF-8 (\DATAAuthorsRmNondecodable/ author names removed), are email addresses instead of names (\DATAAuthorsRmEmail/ ``names''), consist of only blank characters (\DATAAuthorsRmBlank/), contain more than 10\% non-letters (\DATAAuthorsRmNonletter/), are longer than 100 characters (\DATAAuthorsRmToolong/).
After filtering, about \DATAAuthorsPlausibleApprox/ authors (\DATAAuthorsPlausiblePct/ of the initial dataset) remained for further analysis.

Note that the amount of public code commits (and authors) contained in the
initial dataset grows exponentially over
time~\cite{swh-provenance-emse}\ifgrowthfig, as shown for commits in
\Cref{fig:growth}\else: from $10^4$ commits in 1971, to $10^6$ in 1998, to
almost $10^9$ in 2020\fi. As a consequence the observed trends tend to be more
stable in recent decades than in 40+ year-old ones, due to statistics taken on
exponentially larger populations.

\paragraph{Geolocation}

\begin{figure}
  \centering
  \includegraphics[clip,trim=6cm 6cm 0 0,width=\linewidth]{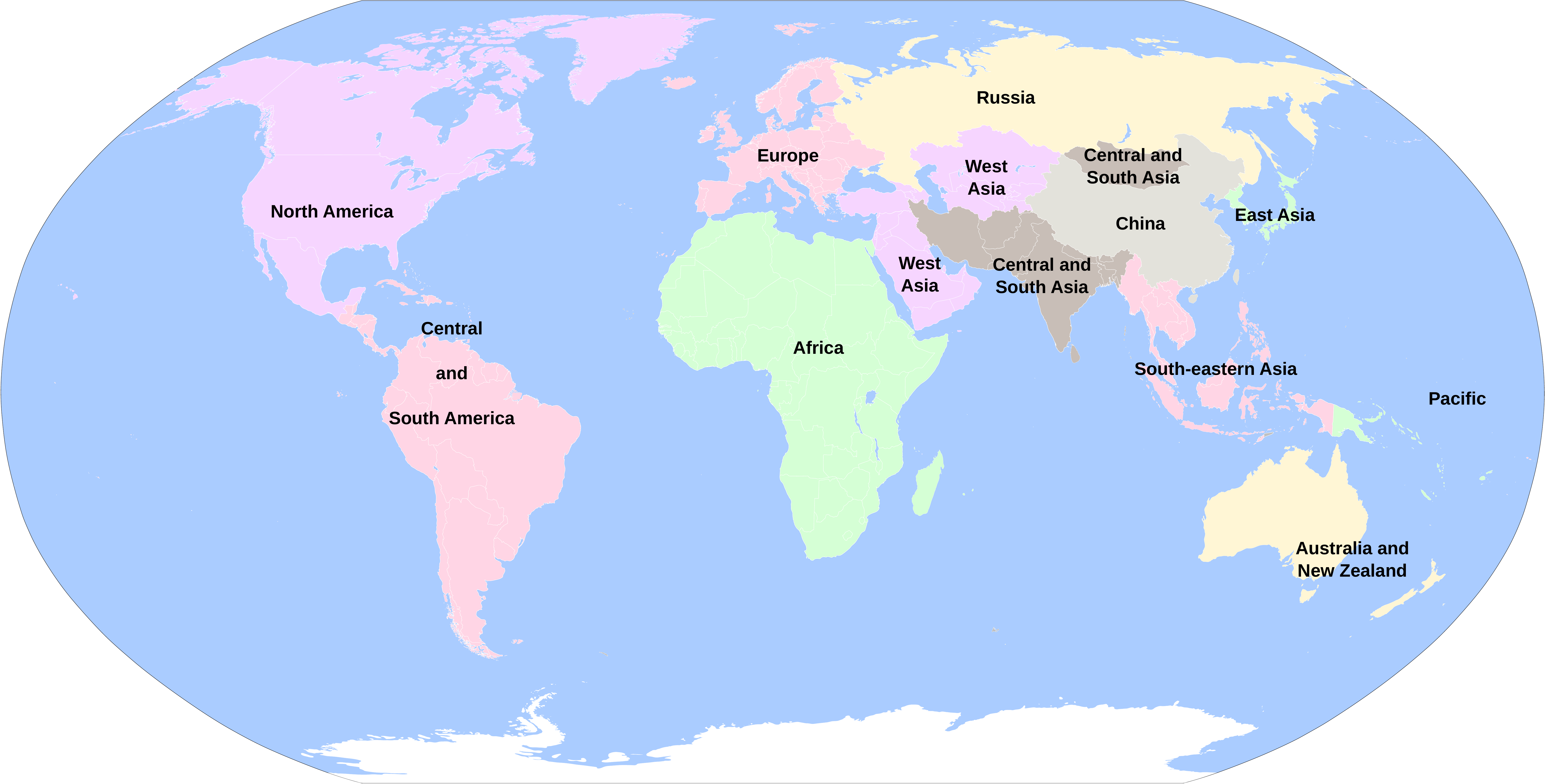}
  \caption{The \DATAWorldRegions/ world regions used as geolocation targets.}
  \label{fig:worldmap}
\end{figure}

As geolocation targets we use macro world regions derived from the United Nations geoscheme~\cite{un1999geoscheme}.
To avoid domination by large countries (e.g., China or Russia) within macro regions, we merged and split some regions based on geographic proximity and the sharing of preeminent cultural identification features, such as spoken language.
\Cref{fig:worldmap} shows the final list of \DATAWorldRegions/ world regions used as geolocation targets in this study.

Geolocation of commit authors to world regions uses the two complementary techniques introduced in~\cite{icse-seis-2022-gender}, briefly recalled below.
The first one relies on the country code top-level domain (ccTLD) of email addresses extracted from commit metadata, e.g., \texttt{.fr}, \texttt{.ru}, \texttt{.cn}, etc.
We started from the IANA list of Latin character ccTLDs~\cite{wikipedia-cctld} and manually mapped each corresponding territory to a target world region.

The second geolocation technique uses the UTC offset of commit timestamps (e.g., UTC-05:00) and author names to determine the most likely world region of the commit author.
For each UTC offset we determine a list of compatible places (country, state, or dependent territory) in the world that, at the time of that commit, had that UTC offset; commit time is key here, as country UTC offsets vary over time due to timezone changes.
To make this determination we use the IANA time zone database~\cite{tzdata}.

Then we assign to each place a score that captures the likelihood that a given author name is characteristic of it.
To this end we use the Forebears dataset of the frequencies of the most common first and family names which, quoting from~\cite{forebear-names}: {\itshape ``provides the approximate incidence of forenames and surnames produced from a database of \num{4 044 546 938} people (55.5\% of living people in 2014). As of September 2019 it covers \num{27 662 801} forenames and \num{27 206 821} surnames in 236 jurisdictions.''}
As in our dataset authors are full name strings (rather than split by first/family name), we first tokenize names (by blanks and case changes) and then lookup individual tokens in both first and family names frequency lists.
For each element found in name lists we multiply the place population\footnotemark{} by the name frequency to obtain a measure that is proportional to the number of persons bearing that name (token) in the specific place.
\footnotetext{To obtain population totals---as the notion of ``place'' is heterogeneous: full countries v.~slices of large countries spanning multiple timezones---we use a mixture of primary sources (e.g., government websites), and non-primary ones (e.g., Wikipedia articles).}
We sum this figure for all elements to obtain a place score, ending up with a list of $\langle$place, score$\rangle$ pairs.
We then partition this list by the world region that a place belongs to and sum the score for all the places in each region to obtain an overall score, corresponding to the likelihood that the commit belongs to a given world region.
We assign the starting commit as coming from the world region with the highest score.

The email-based technique suffers from the limited and unbalanced use of ccTLDs: most developers use generic TLDs such as \texttt{.com}, \texttt{.org}, or \texttt{.net}.
Moreover this does not happen uniformly across zones: US-based developers, for example, use the \texttt{.us} ccTLD much more seldomly than their European counterparts.
On the other hand the offset/name-based technique relies on the UTC offset of the commit timestamps.
Due to tool configurations on developer setups, a large number of commits in the dataset has an UTC offset equal to zero.
This affects less recent commits (\DATACommitsTZZTwoThousandTwenty/ of 2020s commits have a zero offset) than older ones (\DATACommitsTZZTwoThousand/ in 2000).
As a result the offset/name-based technique could end up detecting a large share of older commits as authored by African developers, and to a lesser extent Europeans.

To counter these issues we combine the two geolocation techniques together by applying the offset/name-based techniques to all commits with a non-zero UTC offset, and the email-based on to all other commits.

 \section{Results and Discussion}
\label{sec:results}

\begin{figure*}
  \centering
  \includegraphics[width=\linewidth]{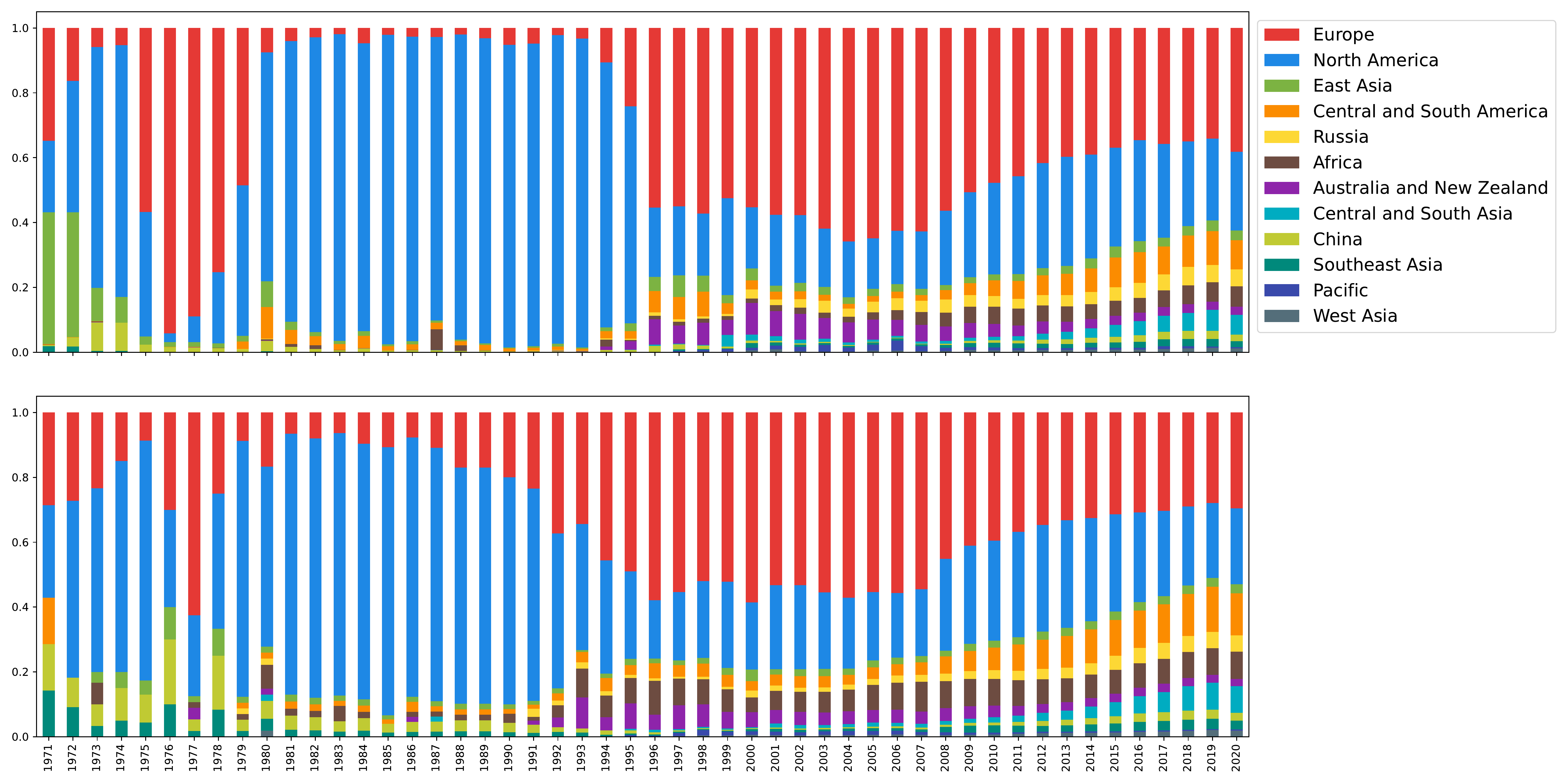}
  \caption{Ratio of commits (above) and active authors (below) by world zone over the 1971--2020 period.}
  \Description[Chart]{Stacked bar chart showing the world zone ratios for commits and authors over the 1971--2020 period.}
  \label{fig:results}
\end{figure*}

To answer \cref{rq:geodiversity} we gathered the number of commits and distinct authors per year and per world zone.
We present the obtained results in \Cref{fig:results} as two stacked bar charts, showing yearly breakdowns for commits and authors respectively.
Every bar represents a year and is partitioned in slices showing the commit/author ratio for each of the world regions of \Cref{fig:worldmap} in that year.
To avoid outliers due to sporadic contributors, in the author chart we only consider authors having contributed at least 5 commits in a given year.

While observing trends in the charts remember that the total numbers of commits and authors grow exponentially over time.
Hence for the first years in the charts, the number of data points in some world regions can be extremely small, with negative consequences on the stability of trends.

\paragraph{Geographic diversity over time}

Overall, the general trend appears to be that the \textbf{geographic diversity in public code is increasing}: North America and Europe alternated their ``dominance'' until the middle of the 90s; from that moment on most other world regions show a slow but steady increment.
This trend of increased participation into public code development includes Central and South Asia (comprising India), Russia, Africa, Central and South America,
Notice that also zones that do not seem to follow this trend, such as Australia and New Zealand, are also increasing their participation, but at a lower speed with respect to other zones.
For example, Australia and New Zealand incremented the absolute number of their commits by about 3 orders of magnitude from 2000 to present days.

Another interesting phenomenon that can be appreciated in both charts is the sudden contraction of contributions from North America in 1995; since the charts depict ratios, this corresponds to other zones, and Europe in particular, increasing their share.
An analysis of the main contributions in the years right before the contraction shows that nine out of ten have \texttt{ucbvax.Berkeley.EDU} as author email domain, and the tenth is Keith Bostic, one of the leading Unix BSD developers, appearing with email \texttt{bostic}.
No developer with the same email domain appears anymore within the first hundred contributors in 1996.
This shows the relevance that BSD Unix and the Computer Systems Research Group at the University of California at Berkeley had in the history of open source software.
The group was disbanded in 1995, partially as a consequence of the so-called UNIX wars~\cite{kernighan2019unixhistory}, and this contributes significantly---also because of the relatively low amount of public code circulating at the time---to the sudden drop of contributions from North America in subsequent years.
Descendant UNIX operating systems based on BSD, such as OpenBSD, FreeBSD, and NetBSD had smaller relevance to world trends due to (i) the increasing amount of open source code coming from elsewhere and (ii) their more geographically diverse developer community.

Another time frame in which the ratios for Europe and North America are subject to large, sudden changes is 1975--79.
A preliminary analysis shows that these ratios are erratic due to the very limited number of commits in those time period, but we were unable to detect a specific root cause.
Trends for those years should be subject to further studies, in collaboration with software historians.

\paragraph{Colonialism}

Another trend that stands out from the charts is that Africa appears to be well represented.
To assess if this results from a methodological bias, we double-checked the commits detected as originating from Africa for timezones included in the $[0, 3]$ range using both the email- the offset/name-based methods.
The results show that the offset/name-based approach assigns 22.7\% of the commits to Africa whereas the email-based one only assigns 2.7\% of them.
While a deeper investigation is in order, it is our opinion that the phenomenon we are witnessing here is a consequence of colonialism, specifically the adoption of Europeans names in African countries.
For example the name Eric, derived from Old Norse, is more popular in Ghana than it is in France or in the UK.
This challenges the ability of the offset/name-based method to correctly differentiate between candidate places.
Together with the fact that several African countries are largely populated, the offset/name-based method could detect European names as originating from Africa.
While this cuts both way, the likelihood of a random person contributing to public code is very different between European countries, all having a well-developed software industry, and African countries that do not all share this trait.

\paragraph{Immigration/emigration}

Another area where a similar phenomenon could be at play is the evolution of Central and South America.
Contribution from this macro region appears to be growing steadily.
To assess if this is the result of a bias introduced by the name-based detection we analyzed the evolution of offset/name-based assignment over time for authors whose email domain is among the top-ten US-based entities in terms of overall contributions (estimated in turn by analyzing the most frequent email domains and manually selecting those belonging to US-based entities).
In 1971 no author with an email from top US-based entities is detected as belonging to Central and South America, whereas in 2019 the ratio is 12\%.
Nowadays more than one tenth of the people email-associated to top US-based entities have popular Central and South American names, which we posit as a likely consequence of immigration into US (emigration from Central and South America).
Since immigration has a much longer history than what we are studying here, what we are witnessing probably includes long-term consequences of it, such as second and third generation immigrants employed in white-collar jobs, such as software development.

 \section{Limitations and Future Work}
\label{sec:conclusion}

We have performed an exploratory, yet very large scale, empirical study of the geographic diversity in public code commits over time.
We have analyzed \CommitCount public commits covering the \DATAYearRange/ time period.
We have geolocated developers to \DATAWorldRegions/ world regions using as signals email domains, timezone offsets, and author names.
Our findings show that the geographic diversity in public code is increasing over time, and markedly so over the past 20--25 years.
Observed trends also co-occur with historical events and macro phenomena like the end of the UNIX wars, increase of coding literacy around the world, colonialism, and immigration.

\medskip
\emph{Limitations.}
This study relies on a combination of two geolocation methods: one based on email domains, another based on commit UTC offsets and author names.
We discussed some of the limitations of either method in \Cref{sec:method}, motivating our decision of restricting the use of the email-based method to commits with a zero UTC offset.
As a consequence, for most commits in the dataset the offset/name-based method is used.
With such method, the frequencies of forenames and surnames are used to rank candidate zones that have a compatible UTC offset at commit time.

A practical consequence of this is that for commits with, say, offset UTC+09:00 the candidate places can be Russia, Japan and Australia, depending on the specific date due to daylight saving time.
Popular forenames and surnames in these regions tend to be quite different so the likelihood of the method to provide a reliable detection is high.
For other offsets the set of popular forenames and surnames from candidate zones can exhibit more substantial overlaps, negatively impacting detection accuracy.
We have discussed some of these cases in \Cref{sec:results}, but other might be lingering in the results impacting observed trends.

The choice of using the email-based method for commits with zero UTC offset, and the offset/name-based method elsewhere, has allowed us to study all developers not having a country-specific email domain (ccTLD), but comes with the risk of under-representing the world zones that have (in part and in some times of the year) an actual UTC offset of zero.

A potential bias in this study could be introduced by the fact that the name database used for offset/name-based geolocation only contains names formed using Latin alphabet characters.
We looked for names containing Chinese, Japanese, and Korean characters in the original dataset, finding only a negligible amount of authors who use non-Latin characters in their VCS names, which leads us to believe that the impact of this issue is minimal.

We did not apply identity merging (e.g., using state-of-the-art tools like SortingHat~\cite{moreno2019sortinghat}), but we do not expect this to be a significant issue because: (a) to introduce bias in author trends the distribution of identity merges around the world should be uneven, which seems unlikely; and (b) the observed commit trends (which would be unaffected by identity merging) are very similar to observed author trends.

We did not systematically remove known bot accounts~\cite{lebeuf2018swbots} from the author dataset, but we did check for the presence of software bots among the top committers of each year. We only found limited traces of continuous integration (CI) bots, used primarily to automate merge commits. After removing CI bots from the dataset the observed global trends were unchanged, therefore this paper presents unfiltered data.

\medskip
\emph{Future work.}
To some extent the above limitations are the price to pay to study such a large dataset: there exists a trade-off between large-scale analysis and accuracy.
We plan nonetheless to further investigate and mitigate them in future work.
Multi-method approaches, merging data mining with social science methods, could be applied to address some of the questions raised in this exploratory study.
While they do not scale to the whole dataset, multi-methods can be adopted to dig deeper into specific aspects, specifically those related to social phenomena.
Software is a social artifact, it is no wonder that aspects related to sociocultural evolution emerge when analyzing its evolution at this scale.

\clearpage


\end{document}